\documentclass[aps,prl,twocolumn,superscriptaddress,showpacs,amsmath,amssymb, amsfonts,10pt]{revtex4-2}
\usepackage{graphicx, placeins}
\usepackage[unicode=true,pdfusetitle,bookmarks=true,bookmarksnumbered=true,bookmarksopen=false, breaklinks=false,pdfborder={0 0 0},pdfborderstyle={},backref=false,colorlinks=false, pdftitle={Superconductivity induced by strong electron-exciton coupling in doped atomically thin semiconductor heterostructures}]{hyperref}
\usepackage{amsmath}
\usepackage{color}
\usepackage{verbatim}
\usepackage[capitalise]{cleveref}
\usepackage[normalem]{ulem} 

\newcommand{\nnl}{\nonumber \\}

\newcommand{\ced}{\hat c^\dagger}
\newcommand{\ce}{\hat{c}^{\phantom{\dagger}}}
\newcommand{\Xd}{\hat X^\dagger}
\newcommand{\X}{\hat{X}^{\phantom{\dagger}}}

\newcommand{\pv}{\mathbf{p}}
\newcommand{\vecp}{\mathbf{p}}
\newcommand{\veck}{\mathbf{k}}
\newcommand{\vecq}{\mathbf{q}}

\newcommand{\kv}{\mathbf{k}}
\newcommand{\qv}{\mathbf{q}}

\usepackage{braket}
\usepackage{appendix}
\usepackage{ulem}
\usepackage{lipsum}  
\usepackage{enumerate} 
\DeclareMathOperator{\STr}{STr}
\usepackage{tikz}
\usepackage{tikz-feynman}
\usepackage{tikz-feynhand}
\usepackage{adjustbox}
\usepackage{cancel}

\usepackage{float}

\begin{document}

\title{Superconductivity induced by strong electron-exciton coupling in doped atomically thin semiconductor heterostructures}

\author{Jonas von Milczewski}
\email[E-Mail: ]{jvmilczewski@mpq.mpg.de}
\affiliation{Institute for Theoretical Physics, Heidelberg University, Philosophenweg 16, 69120 Heidelberg, Germany}
\affiliation{Max-Planck-Institute of Quantum Optics, Hans-Kopfermann-Strasse 1, 85748 Garching, Germany}
\affiliation{Department of Physics, Harvard University, Cambridge, Massachusetts 02138, USA}
\author{Xin Chen}
\affiliation{Institute for Theoretical Physics, Heidelberg University, Philosophenweg 16, 69120 Heidelberg, Germany}
\author{Atac Imamoglu}
\affiliation{Institute of Quantum Electronics, ETH Zurich, CH-8093 Zurich, Switzerland}
\author{Richard Schmidt}
\affiliation{Institute for Theoretical Physics, Heidelberg University, Philosophenweg 16, 69120 Heidelberg, Germany}

\date{\today}
\begin{abstract}
We study a mechanism to induce superconductivity in atomically thin semiconductors  where excitons mediate an effective attraction between electrons. Our model  includes interaction effects beyond the paradigm  of phonon-mediated superconductivity and connects to the well-established limits of   Bose and Fermi polarons. By  accounting for the strong-coupling physics of trions,  we find that the effective electron-exciton interaction develops a strong frequency and momentum dependence accompanied by the system undergoing an emerging BCS-BEC crossover from weakly bound $s$-wave Cooper pairs to a superfluid of bipolarons. Even at strong-coupling the bipolarons remain relatively light,  resulting in critical temperatures of  up to 10\% of the Fermi temperature. This renders  heterostructures of two-dimensional materials  a promising candidate to realize superconductivity  at high critical temperatures set by electron doping and  trion binding energies.

\end{abstract}

\maketitle 

In the past decade van der Waals (vdW) materials   have been shown to host a plethora of quantum phases of matter ranging from  Mott and Wigner crystals \cite{Regan2020,Shimazaki2021,
Zhou2021,Smolenski2022}, the anomalous quantum Hall effect \cite{Serlin2020,Xie2022,Cai2023,Popert2022}, chiral edge states \cite{Sharpe2019} and Chern insulators \cite{Cao2018} to interaction-driven insulators \cite{Liu2020}. Following  Ising pairing observations in superconducting NbSe2 monolayers
\cite{Xi2015}, the discovery of  
superconductivity in  magic-angle  graphene \cite{Cao2018a,Yankowitz2019} and twisted bilayers of atomically thin semiconductors
\cite{Wang_2020} have advanced vdW-materials  as a platform to realize novel forms of superconductivity.

The existence  of strongly bound excitons in transition metal dichalcogenides (TMD) \cite{Wang2018} has inspired studies exploring  new routes to  superconductivity. Recently, repulsive pairing mechanisms in twisted Moir\'{e} materials came into focus \cite{Crepel_2023,Sun2021,Slagle2020,Crepel2021,Crepel2022,Crepel2022a,He2023}, where flat bands limit Fermi energies and thus the critical transition temperature. Without flat bands, theoretical works have explored exciton-mediated interactions between electrons analogous to  phonon-exchange in conventional BCS theory \cite{Enss2009,Laussy2010,Cotlet2016,Villegas2019,Sun2021b, Kinnunen2018,Julku_2022}. However, TMDs feature an exciton-electron coupling strong enough to feature exciton-electron bound states, trions, that remain  stable at room temperature \cite{Wang2018}, not captured by previously employed Fr\"ohlich-type models \cite{Enss2009,Laussy2010,Cotlet2016, Kinnunen2018,Julku_2022}.  Including this non-perturbative pairing physics in theoretical models  has remained  a central challenge, and begs the question how  trion formation  impacts  superconductivity.

In this letter, we present a theory of boson-induced superconductivity which incorporates the strong-coupling physics of the Bose-Fermi mixtures \cite{Bertaina2013,Ludwig2011,Guidini2015,vonMilczewski2022,Duda2023,tan2022bose} comprised of excitons and electrons. Our work does not rely on flat bands and applies to heterostructures of vdW-materials  where electrons  interact with excitons in separated layers. We account for trion formation by considering beyond-linear electron-exciton coupling terms that extend  the Fröhlich paradigm of electron-phonon exchange \cite{Froehlich1954,Holstein1959}. We find that the effective exciton-electron vertex becomes strongly retarded and non-local leading to strong dressing of electrons by the excitonic background. Tuning the doping level in the TMD, mutual dressing of electrons and excitons leads to an emergent crossover from a weak-coupling BCS superconductor into a superfluid state of bipolarons, akin to the BCS-BEC crossover observed in cold atoms  \cite{Jochim2003,Greiner2003,Zwierlein2003,Regal2004,Zwierlein2004,Chin_2004,Kinast_2004,Bourdel_2004,Strecker_2003,Zwierlein2006,Ku2012,Feld2011,Sommer2012,Ries2015,Murthy2018,Sobirey2021,Holten2022}. Remarkably, we find  bipolarons to remain relatively light, facilitating  transition temperature reaching values of up to 10 \% of the Fermi temperature. The physics of a BCS-BEC crossover emerging from mediated interactions complements the direct interaction mechanism in cold atoms, opening perspectives to reach high  transition temperatures in vdW-materials.

\begin{figure}[t]
\includegraphics[width=\linewidth]{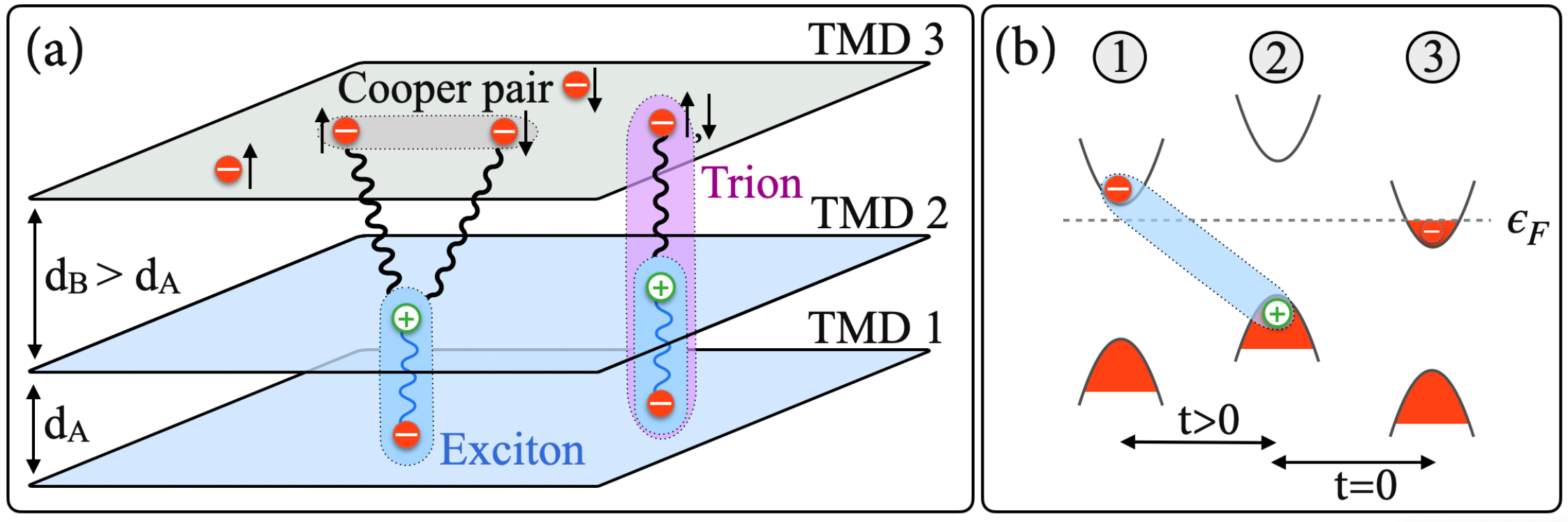}
\caption{(a) Illustration of the TMD heterostructure. Charge carriers in  tunnel-coupled layers  1 and 2 form interlayer excitons which can bind  with electrons in the spatially separated, tunnel-decoupled ($t=0$) top layer into deeply-bound trions.  (b) Using  gates the band alignment of the  layers can be modified to allow doping of the third layer, while the interlayer (12) exciton remains stable. $\epsilon_F$ marks the  Fermi level.}
\label{Illustration}
\end{figure}

\textbf{\textit{Model and Method.---}} We start from a two-dimensional Fermi gas of electrons ($\hat c_{\uparrow\vecp}$, $\hat c_{\downarrow\vecp}$) in absence of a magnetic field. The electrons interact with long-lived interlayer excitons  in a spatially separated heterobilayer which could be realized in a MX$_2$-hBN-M'X$_2$-hBN$_n$-MX$_2$  heterostructure. Here  M$\neq$M' label transition metal, and X chalcogen atoms, and hBN$_{(n)}$ labels one (n) layers of hexagonal boron nitride separating the semiconductor layers as shown in \cref{Illustration}(a).

 Electron tunneling between the top layers is fully suppressed by a large layer separation $d_B>d_A$ enabling  $s$-wave pairing between electrons in the top layer. Gating can be employed to allow for  doping of layer 3  in presence of   long-lived interlayer-(12) excitons. Since interlayer-(12) exciton energies for vanishing separation of the lower TMD layers 1 and 2 would be in the range  $100$ to $150$ meV
\cite{Amelio2023}, Fermi energies  of around $30$ meV in  TMD layer 3 would be possible. Importantly, due to the dipolar character of the system, the
interlayer-(123) trion can have a substantial binding energy $\epsilon_T\sim 30$ meV comparable with the Fermi energy, realizing a strong coupling regime. We also emphasize that the generation of interlayer excitons need not require optical excitation \cite{Shan2021,Wang2022}.
Due to the tunnel decoupling of the top layer, the attractive charge-dipole interaction between electrons  and the interlayer excitons, described by operators $\hat X^\dagger_\vecp$, is independent of spin and can be modelled by an attractive contact interaction of strength $g$ containing the trion energy $\epsilon_T$ \cite{Adhikari1986,Randeria1989}. In experiments the value of $\epsilon_T$ could be tuned 
by changing the thickness of the hBN layer separating TMD layers 2 and 3, or 
using dielectric engineering \cite{Raja2017,Steinleitner2018,Wang2018}.
The corresponding Hamiltonian is given by 
\begin{align}\label{HamiltonianOrg}
    \hat H &= \sum_{\sigma=\uparrow,\downarrow}\sum_{\veck} \epsilon^c_\veck \ced_{\sigma\veck}\ce_{\sigma\veck}+\sum_\veck \epsilon^X_\veck \Xd_{\veck}\X_{\veck}\nonumber\\
    &+ \frac{ g \sqrt{n_0}}{\sqrt{A}}\sum_{\sigma=\uparrow,\downarrow}\sum_{\veck\vecq}\ced_{\sigma\veck+\vecq}\ce_{\sigma\veck}(\Xd_\vecq +\X_{-\vecq})\nonumber\\ &
    + \frac{g}{A} \sum_{\sigma=\uparrow,\downarrow}\sum_{\veck\veck'\vecq}  \ced_{\sigma\veck+\vecq}\ce_{\sigma\veck}\Xd_{\veck'-\vecq} \X_{\veck'},
\end{align}
with $A$ the system area. Assuming an effective mass approximation, the electron and exciton dispersion relations are $\epsilon^c_\kv= \kv^2/ 2m_F$ and $\epsilon^X_\kv= \kv^2/2 m_B$. Although we investigate  superconductivity in TMD, the Hamiltonian in \cref{HamiltonianOrg} may also be realized in ultracold atomic systems where  mass ratios between bosons and fermions can vary substantially. Considering the universal relevance of the model, we work at an equal mass ratio of excitons and electrons $m_B=m_F$. As the Fermi gas is spin-balanced, both components $\sigma=\uparrow,\downarrow$ are described by the  Fermi wavevector  $k_F  = \sqrt{4 \pi n_F}$ with  density $n_F$. The  Fermi level $\epsilon_F$ and temperature $T_F$ are given by $\epsilon_F= T_F=k_F^2/2 m$. We set  $\hbar=k_B=1$.

We employ a mean-field description of the Bose gas that is sufficient to demonstrate the  mechanism of  exciton-induced superconductivity  enhanced by the presence of trions.  This mean-field picture,  an exciton gas  described by a condensate of density $n_0$, is justified by the algebraic decay of the boson correlator in the BKT phase \cite{MWThm,Hohenberg1967,Berezinsky1972,Kosterlitz1973,Petrov2003} on scales larger than the range of induced interactions. In Eq.~\eqref{HamiltonianOrg} we expanded in fluctuations around the condensate, i.e. $\X_\veck\to \delta_{\veck,0} \sqrt{n_0 A}+\X_\veck$.
Considering the  much smaller  separation between layers 1 and 2 compared to recent experiments \cite{Shan2021}, we can consider the regime of a weakly interacting exciton gas  with  healing length  $\xi$  much larger than the interelectron distance. Moreover, considering  that the exciton-electron interaction dominantly probes the particle-like branch of the exciton Bogoliubov dispersion we treat the excitons as an ideal Bose gas.

The first interaction term in Eq.~\eqref{HamiltonianOrg} describes a Fr\"ohlich-type electron-phonon interaction $\lambda{\sim} g\sqrt{n_0}$. In  perturbative approaches to exciton-induced superconductivity \cite{Laussy2010,Cotlet2016,Kinnunen2018,Julku_2022},  induced interactions between electrons originated solely from this term and  scale with $\lambda^2$; i.e. independent of the sign of $\lambda$. However, the microscopic origin of this phonon-like interaction is the attractive potential parametrized by the last term ${\sim} g$ in \cref{HamiltonianOrg}. This term is responsible for trion formation, and its relevance has been demonstrated by  observations in cold atoms and TMD showing strong deviations from the Fr\"ohlich model \cite{Hu2016,jorgensen2016,yan2020bose,tan2022bose}. Using a renormalization group (RG) analysis presented in the Supplemental Materials (SM) \cite{SM}, we  show this term to be RG-relevant and   crucial in the strong-coupling regime. Unlike previous works we  consider this term fully and study its non-perturbative effect on exciton-induced electron pairing. 

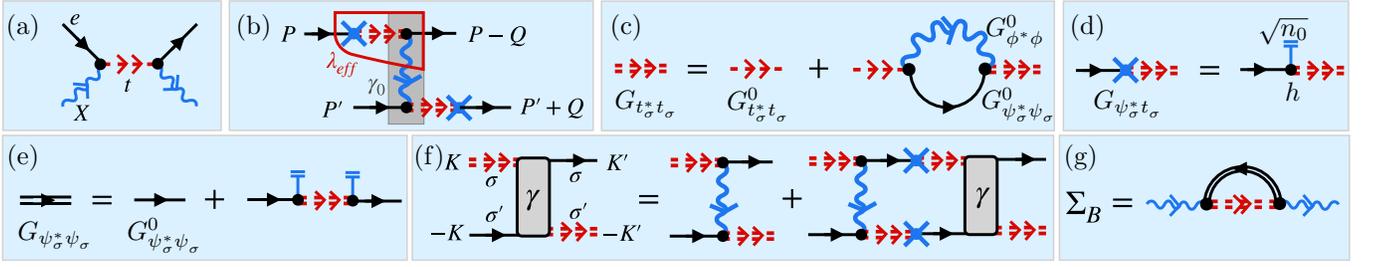
\begin{figure*}[t!]
\normalsize
\begin{tikzpicture}
    \node[anchor=south west,inner sep=0] (image) at (0,0) {\includegraphics[width=\linewidth]{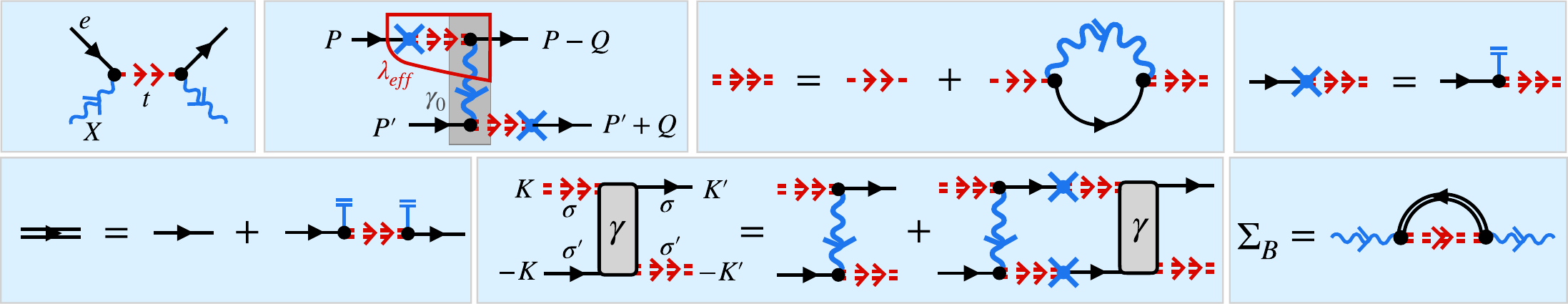}};
    \begin{scope}[x={(image.south east)},y={(image.north west)}]
        \node[align=right] at (0.016,1-0.105) {(a)};
        \node[align=right] at (0.212-.025,1-0.105) {(b)};
         \node[align=right] at (0.503-.04,1-0.105) {(c)};
         \node[align=right] at (0.805,1-0.105) {(d)};
         \node[align=right] at (0.016,1-0.605) {(e)};
         \node[align=right] at (0.317,1-0.605) {(f)};
         \node[align=right] at (0.80,1-0.605) {(g)};
         \node[align=right] at (.95,1-0.12) {$\sqrt{n_0}$};
         \node[align=right] at (.957,1-0.35) {$h$};
         \node[align=right] at (0.043+.435,1-0.9+.5) {$G_{t^*_\sigma t^{\phantom{*}}_\sigma}$};
         \node[align=right] at (0.126+.435,1-0.9+.5) {$G^0_{t^*_\sigma t^{\phantom{*}}_\sigma}$};
         \node[align=right] at (0.316+.435,1-0.62+.5) {$G^0_{\phi^* \phi}$};
        \node[align=right] at (0.319+.435,1-0.9+.5) {$G^0_{\psi^*_\sigma \psi^{\phantom{*}}_\sigma}$};
        \node[align=right] at (0.835,1-0.9+.5) {$G_{\psi^*_\sigma t^{\phantom{*}}_\sigma}$};
        \node[align=right] at (0.04,1-0.9) {$G_{\psi^*_\sigma \psi^{\phantom{*}}_\sigma}$};
        \node[align=right] at (0.121,1-0.9) {$G^0_{\psi^*_\sigma \psi^{\phantom{*}}_\sigma}$};
    \end{scope}
\end{tikzpicture}
\caption{Feynman diagrams for the (a)  bare exciton-electron scattering vertex, (b) induced electron-electron vertex, (c) full trion propagator (red double line), (d) full off-diagonal electron-trion propagator (black, red with cross), (e) full electron propagator (double black line), (f) renormalized electron-trion scattering vertex. The exciton self-energy (g) is used within the Hugenholtz-Pines condition. Bare propagators are shown as single lines and contain an infinite series of condensate insertions (blue double lines), coupling vertices are shown as black dots. See the SM \cite{SM} for analytical expressions.
}
\label{FigFeynCom}
\end{figure*}

To study electron pairing we employ finite-temperature quantum field theory. Using a diagrammatic approach, it is  practical to employ a two-channel model that is  equivalent to Eq.~\eqref{HamiltonianOrg}. To arrive at  this model one employs a Hubbard-Stratonovich transformation where a trion field $t$ manifests the strong-coupling physics and  formally mediates the electron-exciton interaction (\cref{FigFeynCom}(a)) \cite{Holland2001,Timmermans2001,Bruun2004}. The corresponding action is  given by
\begin{align}\label{action}
&S =\int_Q\! \!\Big( \phi^\ast_Q P_{\phi}(Q) \phi_Q 
+\psi_{\sigma,Q}^\ast P_{\psi}(Q)\psi_{\sigma,Q}+ t^{\ast}_{\sigma,Q} P_{t}^{0} (Q) t_{\sigma,Q} \nnl
&+ h \sqrt{n_0}\left[t_{\sigma,Q}^\ast \psi^{\phantom{\ast}}_{\sigma,Q} +\text{h.c.} \right]
\! \! \Big)+ h \!  \!  \int_{P ,Q}\!\!  \!\!  \!
\!  \left[ \psi^{\ast}_{\sigma,Q-P} \phi^{\ast}_P t^{\phantom{\ast}}_{\sigma,Q}+\text{h.c.}\right]\! ,
\end{align}
where  $g =- h^2/ P_t^{0}$ \cite{Lurie1964,Nikolic2007} establishes  equivalence of  models   \eqref{HamiltonianOrg} and \eqref{action} in the contact interaction limit $h\to \infty$ 
 \footnotetext[1]{In this limit, the models   \eqref{HamiltonianOrg} and \eqref{action} are fully equivalent and describe the same physical phenomena. All physical quantities within model   \eqref{HamiltonianOrg} are equivalent to a quantity in model \eqref{action}, that is independent of $h$ in the limit $h\to \infty$. For example, the $T$-matrix in model \eqref{HamiltonianOrg} is equivalent to $T=-h^2  G_{t^*_{\sigma} t^{\phantom{*}}_{\sigma}}$ in model \eqref{action}.
Similarly, $\gamma/h^2$ is independent of $h$. In fact, after taking the limit, $h$ can be eliminated completely and never needs to be specified. Hence, rather than introducing model \eqref{action}, one may also conduct this analysis in terms of model \eqref{HamiltonianOrg}, in that case, however, the corresponding diagrammatics will be more complicated than those described in \cref{FigFeynCom,propagators}.}\footnotemark[1].
The fields $\rho\in\{\psi_{\sigma,Q}, t_{\sigma,Q}, \phi_Q\}$ correspond to  electrons,  trions, and  fluctuations of the exciton gas around its mean value $\sqrt{n_0}$.  Capital letters $Q=(\vecq,\omega_n)$  refer to momenta $\vecq$ and Matsubara frequencies $\omega_n$,  $P_{\phi/\psi}(\qv,\omega)=(- i \omega +\epsilon^{X/c}_\qv -\mu_{B/F})$, and $\int_Q$ contains  Matsubara and spin summation. Electron and exciton chemical potentials  are denoted by $\mu_F, \mu_B$. 

The exciton condensate hybridizes  electrons and trions into a joint excitation (see the term   $ {\sim} \sqrt{n_0} t^*_\sigma \psi^{\phantom{*}}_\sigma$ in~\cref{action}). This hybridization is key for inducing the electron-electron interaction shown in \cref{FigFeynCom}(b). Due to  hybridization, this vertex is internally governed by a trion-electron scattering vertex at tree level $   h^2 \gamma^{0}(Q) t^\ast_\sigma \psi_{\sigma'}^\ast \psi^{\phantom{\ast}}_\sigma t^{\phantom{\ast}}_{\sigma'}$ 
(gray box in \cref{FigFeynCom}(b)), where $\gamma^{0}(Q)=1/P_\phi (Q)$ represents the exchange of an exciton. We  study  exciton-induced Cooper pair formation  in terms of the renormalization of this trion-electron vertex, accounting for the infinite ladder of exciton exchanges (\cref{FigFeynCom}(f)). In this ladder resummation, the strong-coupling physics between excitons  and electrons  is accounted for by  the self-energy  $\Sigma^t_\sigma(\pv,\omega)$ of the trion field  (Fig.~\ref{FigFeynCom}(c)).  As a result of the $(\pv,\omega)$-dependence of $\Sigma_t$, all vertices involving the exchange of a dressed trion (double red line in \cref{FigFeynCom}) become retarded and non-local. In particular, this applies to the effective electron-exciton vertex $\lambda_{\text{eff}}$ (red box in \cref{FigFeynCom}(b)), which appears as a main building block of the  diagrammatic resummation in \cref{FigFeynCom}(f) capturing the superconducting instability, adding  a new ingredient to the mechanism of exciton-induced superconductivity.

We approach the pairing problem within a non-self-consistent $T$-matrix (NSCT) approach \cite{Chevy2006,Combescot2007,Punk_2009,Schmidt2011,Trefzger_2012,Zoellner2011,Schmidt2012,vonMilczewski2022} (for details see SM \cite{SM}), describing both the non-perturbative electron-exciton scattering physics, and the self-energy corrections for the excitons and electrons via the diagrams shown in Fig.~\ref{FigFeynCom}(d,e,g). In this way we recover the associated Fermi  \cite{Zoellner2011,Parish2011,Schmidt2012,Bertaina2012,Parish2013,Kroiss2014,Vlietinck2014,vonMilczewski2022} and Bose polaron formation \cite{rath2013,Isaule2021}   observed in ultracold atoms  \cite{Hu2016,jorgensen2016,yan2020bose,Kohstall_2012,Schirotzek2009,Koschorreck2012,Ness2020,Fritsche2021,Duda2023} and TMDs \cite{sidler2017fermi,Goldstein2020,Xiao2021,Liu2021,tan2022bose,Zipfel2022}. Recently  this approach was shown to apply equally to nearly population balanced, strongly-coupled Bose-Fermi mixtures \cite{Ludwig2011,Guidini2015,vonMilczewski2022,Duda2023}. 
Hence, our approach is  based on a  model \eqref{HamiltonianOrg}  that has been firmly tested in  experiments.

We incorporate  self-energy effects  by using the renormalized (matrix-valued) Green's function $G$, 
\begin{align}
    \left(G^{-1}\right)_{\rho \rho'}=  \left(G_{0}^{-1}\right)_{\rho \rho'}-\frac{\partial^2}{\partial \rho \partial \rho'} \! \sum_\sigma t^*_\sigma \Sigma^t_\sigma t^{\phantom{*}}_\sigma\Big|_{\psi_\sigma, \phi,t_\sigma=0}, \label{propagators}
\end{align}
rather than the bare  Green's  function $G_{0}$  
$\left(G_{0}^{-1}\right)_{\rho \rho'}= \frac{\partial^2}{\partial \rho \partial \rho'} S |_{(\psi_\sigma, \phi,t_\sigma)=0}$.
 In \cref{propagators}  we  suppressed $(\pv,\omega)$-arguments; for  analytic expressions see \cite{SM}. The pole of the trion Green's function $G_{t^*_{\sigma} t^{\phantom{*}}_\sigma}$ in the two-body  limit determines the trion energy $\epsilon_T$ \cite{Randeria1989}.

The electron pairing problem is solved using a Bethe-Salpether equation  for the renormalized electron-trion vertex  function (\cref{FigFeynCom}(f)), 
\begin{align}\label{BSE}
     \gamma(K&-K’,K,K’)_{\sigma,\sigma’}=\gamma_{0}(K-K’) 
    + h^2 \int_P \! \!  \big[\gamma_{0}(K-P) \nnl
    & \times G_{\psi^*_{\sigma}t^{\phantom{*}}_{\sigma}}(P)  G_{t^{*}_{\sigma’}\psi^{\phantom{*}}_{\sigma’}}(-P)  \gamma(P-K’,P,K’)_{\sigma,\sigma’}\big].
\end{align}
Within  model \eqref{HamiltonianOrg}, the physical quantity corresponding to $\gamma$ has form $\gamma/h^2$ (which for $h\to \infty$ has no $h$-dependence) and a singularity in $\gamma/h^2$ indicates a pairing instability \footnotemark[1].  As Pauli exclusion suppresses bound state formation between equal spin fermions, we consider the $s$-wave projection of \cref{BSE}, denoted by $\tilde{\gamma}$, to study $s$-wave pairing of opposite spin electrons, $\sigma\neq\sigma'$ 
\footnotetext[2]{We assume equal interaction strength of $\uparrow$-, $\downarrow$-electrons with the excitons as the layer separation strongly suppresses exchange effects.}\footnotemark[2].
In Eq.~\eqref{BSE} we focus on a subset of   diagrams where the  $\gamma$-vertex couples to itself and which contains 
the off-diagonal Green's function $G_{\psi^*_{\sigma}t^{\phantom{*}}_{\sigma}}$ (Fig.~\ref{FigFeynCom}(d)). This disregards diagrams originating purely from the exchange of thermal excitons as at  low temperatures   the majority of bosons condenses and thermal excitations  have subleading contributions. Furthermore, this approximation leaves out  exchange diagrams leading to bosonic three-body bound state formation already in the few-body limit \cite{Kartavtsev2007}. Hence we expect that including such diagrams would   enhance Cooper pair formation  even further.

For  given  $n_0$,  $n_F$,   $T$ and $\epsilon_T$, the chemical potentials $\mu_F$ and $\mu_B$ contained in \cref{propagators}  are determined self-consistently to fulfill two conditions:
\begin{enumerate}[(i)]
\item The number equation to set the density of fermions, $n_{F}=\frac{T}{(2\pi)^2} \int d \pv\sum_n G_{\psi^*_\sigma \psi^{\phantom{*}}_\sigma} (\pv, \omega_n)$.
\item The Hugenholtz-Pines relation  $0=\mu_B+ \Sigma_B(\mathbf{0},0)$ to ensure that  excitations from the 
condensate are gapless (the condensate is kept as a background field in our model). The boson self-energy $\Sigma_B(\vecp,\omega)$ (Fig.~\ref{FigFeynCom}(g)) is given in the SM \cite{SM}.
\end{enumerate}

These two conditions naturally incorporate the physics of both Bose and Fermi polarons (see SM \cite{SM}): For a vanishing fermion density $n_{F}=0$ (see \cref{FigFeynCom}(e)), (i) determines the  energy of Bose polarons \cite{rath2013} in agreement with experiments \cite{Hu2016,jorgensen2016,yan2020bose,tan2022bose}.  For vanishing boson density $n_0=0$, (ii)  yields the Fermi polaron energy  in excellent agreement with experiments \cite{Chevy2006,Combescot2007,Punk_2009,Zoellner2011,Schmidt2012,Parish2011,Trefzger_2012,Schirotzek2009,Koschorreck2012,sidler2017fermi,Ness2020,Fritsche2021,Goldstein2020,Xiao2021,Liu2021,Zipfel2022}.

\begin{figure}[t]
\includegraphics[width=\linewidth]{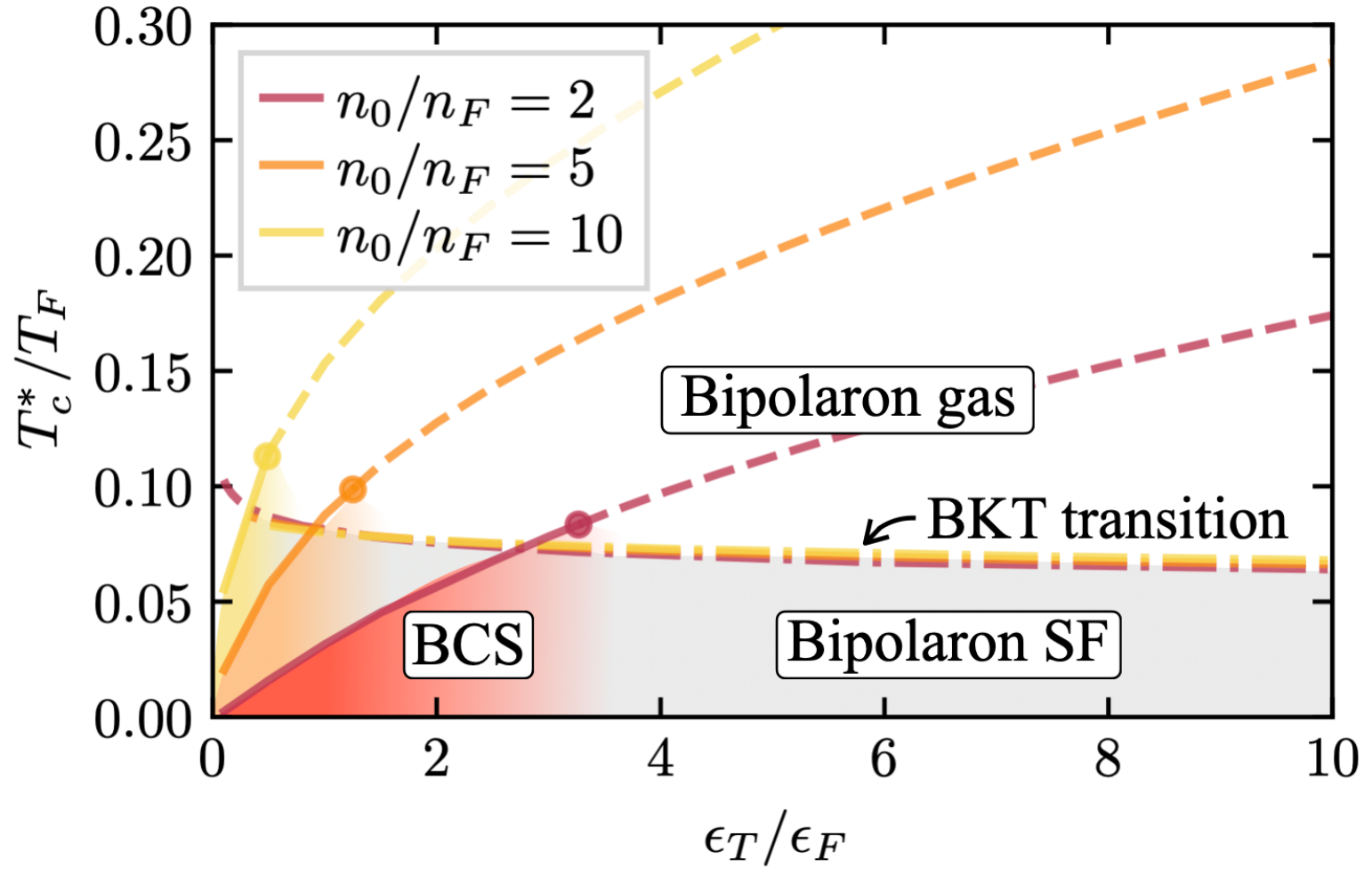}
\caption{Critical pairing temperatures $T_c^*/T_F$ as  function of trion energy  $\epsilon_T/\epsilon_F$ for various condensate densities. In the BCS limit where $T_c^*$ is close to the BKT temperature $T_c$  (see \cref{bcsboundary}) the data is shown as solid lines, while at strong coupling bipolarons are formed and  dashed lines represent their dissociation temperature.  Dash-dotted lines give $T_c=T_{BKT}$ (\cref{BKTtemp})
where the bipolaron gas turns superfluid. 
 }
\label{crittemp}
\end{figure}

\textbf{\textit{Critical pairing temperature.---}} The critical temperature $T_c^*$ for instability towards $s$-wave pairing is determined by lowering the temperature until  $\tilde{\gamma}/h^2$ diverges. Results for $T_c^*$ are shown in \cref{crittemp} in dependence of the dimensionless trion energy $\epsilon_T/\epsilon_F$  for different exciton densities $n_0/\epsilon_F$. 

As $\epsilon_T/\epsilon_F$ is increased,  $T_c^*/T_F$ increases monotonously. Similarly, $T_c^*$  increases with $n_0$,  emphasizing excitons as the mediators of interactions. Increasing interactions and condensate density  dresses  bosons and fermions by many-body fluctuations, strongly increasing boson and fermion chemical potentials (see \cite{SM}) as imposed by conditions (i) and (ii).  Since the chemical potentials enter the propagators in our  diagrammatics, they suppress pairing fluctuations. Despite this suppression,  we find $T_c^{*}$ to increase without apparent bound.

 In the weak-coupling limit, where $T_c^*$ is small, an effective BCS theory applies. In the BCS regime, it has been established that $T_c^*$ is  close to the actual BKT transition temperature $T_c$ towards superfluidity  \cite{Miyake1983,Petrov2003,Chubukov2013}. This equivalence typically applies when the size of Cooper pairs $l_C$ is extended over many interfermion distances $d{\sim} k_F^{-1}$. However, as $l_C$ becomes comparable to the interfermion distance, $T_c^*$ rather starts to indicate  only the formation of pairs but  not their transition into a superfluid state, i.e. $T_c < T_c^*$. 
 
\begin{figure}[t]
\includegraphics[width=0.95\linewidth]{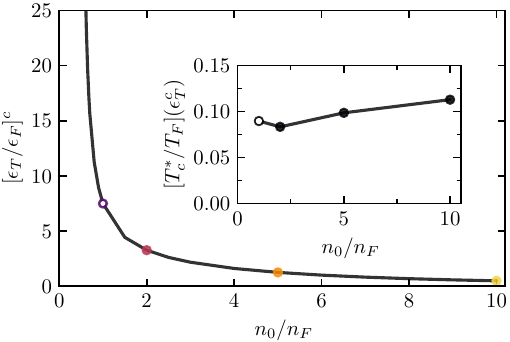}
\caption{Critical interaction strength $\epsilon_T/\epsilon_F$ (as  function of condensate density $n_0/n_F$) beyond which the bipolaron binding energy exceeds the Fermi energy. Colored points mark densities $n_0/n_F$ 
shown  in \cref{crittemp}.  Inset: Critical temperature at the critical interaction strengths. 
For illustration we also show a data point for $n_0/n_F=1$ (open circle) 
where trion formation might deplete the condensate to a degree beyond our description \cite{Parish2011,Bertaina2012,Guidini2015,Duda2023,vonMilczewski2022}. 
}\label{bcsboundary}
\end{figure}

 At strong coupling a different criterion to determine $T_c$ is thus required. At zero temperature the vertex described in \cref{BSE} admits a bound state between two electrons even in the polaron limit where $n_F=0$, $n_0>0$ \cite{Camacho2018}, representing a bipolaron. By  determining where the bipolaron  energy $E_{BP}$ becomes comparable to the Fermi energy $\epsilon_F$, we  estimate  where the Cooper pair size becomes comparable to the interparticle distance, $l_C\approx d$.  To this end, we calculate  $E_{BP}$ by solving \cref{BSE} in the polaron limit. The corresponding critical values  $[\epsilon_T/\epsilon_F]^c$  are shown in \cref{crittemp} as colored dots and the full dependence on $n_0/n_F$ is shown in \cref{bcsboundary}. Beyond  $[\epsilon_T/\epsilon_F]^c$ a description in terms of pairs that immediately condense as they form is clearly  invalid. In this regime, $T_c^*$ should instead be  regarded as the molecular dissociation temperature of bipolarons forming a thermal bipolaron gas that has to be cooled further to facilitate  transition into a superfluid state.

For large $\epsilon_T/\epsilon_F$, bipolarons are sufficiently deeply bound that,   at finite fermion density $n_F>0$, the system can be described by an effective theory of weakly interacting, rigid bosons using BKT theory \cite{Fisher1988,Proko2001,Proko2002,Petrov2003}. To estimate the critical temperature for the BKT transition into the  superfluid state, we employ  the Nelson criterion \cite{Fisher1988,Proko2001,Proko2002,Petrov2003},
\begin{align}
    T_{BKT} = \frac{2 \pi n_F}{m_{BP}} \frac{1}{ \log \left(\frac{\eta}{4 \pi} \log \left(\frac{1}{n_F d_*^2}\right) \right)}. \label{BKTtemp}
\end{align}
with density of bipolarons  $n_F$, and $\eta \approx 380$ \cite{Petrov2003}. The   bipolaron-bipolaron scattering length  $d_*$ and  effective bipolaron effective  mass $m_{BP}$ are  computed in the SM \cite{SM}. We note that  bipolarons remain relatively light which, similar to  recent studies of bipolarons in  the Peierls model \cite{Sous2018,Carbone2021}, facilitates rather large values of $T_c$. The BKT transition temperatures obtained from \cref{BKTtemp} are shown in \cref{crittemp} as dashed-dotted lines and  predictions from the BCS limit and the bipolaron theory intersect in the expected region indicated by  dots in \cref{crittemp}.

Connecting the $T_c$ results  from weak to strong coupling makes evident that the systems is governed by  an \textit{emerging} BCS-BEC crossover from superfluid Cooper pairs  (`BCS' in \cref{crittemp}) to a quasi-condensate of bipolarons (`Bipolaron SF' in \cref{crittemp}).

Despite originating from mediated interactions, the maximal $T_c/T_F$   reaches values on the order of $10\%$, not far below  values obtained in the conventional model of the  BCS-BEC crossover \cite{Eagles1969,Leggett1980,Miyake1983,Nozieres1985,Randeria1989,Randeria1990,SchmittRink1989,Drechsler1992,Petrov2003,Botelho2006,Levinsen2015}  describing fermions that interact via direct, short-range  potentials. We estimate this maximum value of $T_c/T_F$ by considering the temperature at the endpoints calculated in  \cref{crittemp}. The results are shown in the inset of \cref{bcsboundary} and demonstrate  insensitivity with respect to the  density of the exciton gas. In particular at exciton densities $n_0/n_F\gg 1$ the critical temperature remains robust. At such densities, neither thermal nor interaction-driven depletion of the condensate ---not considered here--- plays a significant role,  attesting to the robustness of the mechanism of trion-enhanced, exciton mediated superconductivity.

\textbf{\textit{Conclusion.---}} Incorporating the strong-coupling physics of  exciton-electron mixtures, we showed that exciton-mediated pairing of electrons in doped, atomically thin semiconductor heterostructures offers a promising route towards realizing superconductivity at high temperatures $T_c/T_F$. 
Our work applies in  experimentally realizable regime where  exciton  densities are larger than electron densities. A unified description of the strong-coupling regime where all scales, $\epsilon_F $, $n_0$, $\epsilon_T$ are of the same order is an interesting venue for future studies. Here, a fully self-consistent treatment of  quasiparticles is required and interaction-driven  condensate depletion may have a significant effect.

We did not discuss the impact of the underlying repulsive Coulomb interaction. While this can be justified by  screening at sufficient electron densities (as evidenced by the agreement of  model \eqref{HamiltonianOrg} with experimental observations \cite{sidler2017fermi}), it remains an open problem to formally study the interplay of Coulomb screening and pairing fluctuations. Ultimately this competition may result in $p$-wave pairing becoming the leading instability in certain density regimes \cite{Li2023} while, in turn, higher-order correlation functions \cite{Kartavtsev2007} may favor the $s$-wave pairing studied in this work. 

\textbf{\textit{Note added.}---}Zerba \textit{et al.} explored a complementary scheme using Feshbach resonances for inducing p-wave superconductivity in Bose-Fermi mixtures realized with TMD heterostructures~\cite{Zerba2023}.

\textbf{\textit{Acknowledgements.}---}
We thank Eugene Demler, Nigel Cooper and Verena Köder for inspiring discussions. This work was supported by the Deutsche Forschungsgemeinschaft under Germany's Excellence Strategy EXC 2181/1 - 390900948 (the Heidelberg STRUCTURES Excellence Cluster). The work of A.I. was supported by the Swiss National Science Foundation (SNSF) under Grant Number 200021-204076.  J.v.M. is also supported by a fellowship of the International Max Planck Research School for Quantum Science and Technology (IMPRS-QST).

\clearpage
\newpage
\begin{widetext}
\begin{center}
\textbf{\large Supplemental Material for `Superconductivity induced by strong electron-exciton coupling in doped atomically thin semiconductors'}\label{Sec:Supp}
\end{center}

\begin{center}
Jonas von Milczewski, Xin Chen, Ata\c{c} \.{I}mamo\u{g}lu, and Richard Schmidt 
\end{center}

\setcounter{equation}{0}
\setcounter{figure}{0}
\setcounter{table}{0}
\setcounter{page}{1}
\makeatletter
\renewcommand{\theequation}{S\arabic{equation}}
\renewcommand{\thefigure}{S\arabic{figure}}
\renewcommand{\bibnumfmt}[1]{[S#1]}

In this supplemental material, we discuss details of the calculations and analysis that led to the results presented in the main text. Throughout this supplemental material we work in units where $ m=1/2$. 
Note, that in principle the action in \cref{action} involves terms of the form $gn_0$. However, $g$ is regulated using an upper momentum cutoff $\Lambda$ \cite{Zoellner2011} and thus these terms vanish as $\Lambda$ is increased and are hence left out in $G_0$ and \cref{action} of the main text.

\section{Renormalization group analysis of the extended Fr\"ohlich model in the few-body limit} 

We conduct a renormalization group (RG) analysis in the few-body limit of the running coupling constants within the Hamiltonian given by \cref{HamiltonianOrg} in the main text. To begin, we consider the action
\begin{align}
S &=\int_Q \phi^\ast_Q P_{\phi}(Q) \phi_Q+\psi_{Q}^\ast P_\psi (Q)\psi_{Q} + \lambda \int_{Q,P} \psi^*_Q \psi^{\phantom{*}}_P \left( \phi^*_{P-Q} + \phi^{\phantom{*}}_{Q-P}\right) + g \int_{Q,P, P'} \psi^*_P \psi^{\phantom{*}}_{Q-P}  \phi^*_{P'}  \phi^{\phantom{*}}_{Q-P'}. \label{actionFRG}
\end{align}
The $\lambda$ term  originates from the term in Eq.~(1) of the main text which is proportional to a condensate density $n_0$, $\lambda {\sim} \sqrt{n_0}g$. We will use a functional RG approach in the following \cite{Wetterich1993}. The truncation of the relevant flowing effective action corresponding to \cref{actionFRG} is given by
\begin{align}\label{truncation}
\Gamma_k =\int_Q \phi^\ast_Q P_{\phi}(Q) \phi_Q+\psi_{Q}^\ast P_\psi (Q)\psi_{Q} + \lambda_k \int_{Q,P} \psi^*_Q \psi^{\phantom{*}}_P \left( \phi^*_{P-Q} + \phi^{\phantom{*}}_{Q-P}\right) + g_k \int_{Q,P, P'} \psi^*_P \psi^{\phantom{*}}_{Q-P}  \phi^*_{P'}  \phi^{\phantom{*}}_{Q-P'}.
\end{align}
Here $k$ is the RG scale, above which all fluctuations have been integrated out. It runs from the UV cutoff scale $k=\Lambda$ to  the infrared at $k=0$. As before, $\psi$ denotes the electron (fermion) field, while $\phi$ denotes the exciton (boson) field.   We fix the initial conditions such that $\lambda_{k=\Lambda}= \lambda$ and $g_{k=\Lambda}= g$. In the following, we treat these running couplings as independent to establish a complete picture of the RG flow of the model. We disregard that the flowing coupling constants may acquire a frequency and momentum dependence during the RG flow and instead use a projection
\begin{align}
    \lambda_k &= \left. \frac{\delta^3}{\delta \phi^*_{0}\delta \psi^{\phantom{*}}_0  \delta  \psi^*_0 
    } \Gamma_k \right|_{\psi=\phi=0}\\
      g_k &= \left. \frac{\delta^4}{\delta \phi^{\phantom{*}}_{0} \delta \phi^*_{0}\delta \psi^{\phantom{*}}_0  \delta  \psi^*_0 
    } \Gamma_k \right|_{\psi=\phi=0},
\end{align}
where the subindices on the fields indicate a projection onto zero frequency and momentum. 
 Using the Wetterich equation \cite{Wetterich1993} we compute the flow of the effective action $\Gamma_k$ 
 \begin{align}
 \partial_k \Gamma_k = \frac{1}{2} \STr \Big[ \Big( \Gamma_k^{(2)}+ R_k\Big)^{-1}\partial_k R_k \Big]  , 
\end{align}
 from which we can determine the flow of the coupling constants $\lambda_k$ and  $g_k$. 
 
 The corresponding diagrams are shown in \cref{FeynmanFlow3}. Choosing a sharp momentum regulator as done in Refs. \cite{Schmidt2011,vonMilczewski2022}, the RG flows are given by 
\begin{align}
    \partial _k g_k &= \tilde{\partial}_k \int_P \left( -\frac{g_k^2}{P_\psi(P)} + \frac{3 \lambda_k^2 g_k}{P_{\psi}(P)^2}- \frac{2 \lambda_k^4}{P_\psi (P)^3}\right) \left(\frac{1}{P_{\phi}(P)}+ \frac{1}{P_{\phi}(-P)}\right) \Theta(|\pv|-k)\label{gfloweq} \\
    \partial _k \lambda_k&= \tilde{\partial}_k \int_P \left( -\frac{g_k \lambda_k}{P_\psi(P)} + \frac{ \lambda_k^3}{P_{\psi}(P)^2}\right) \left(\frac{1}{P_{\phi}(P)}+ \frac{1}{P_{\phi}(-P)}\right) \Theta(|\pv|-k).\label{lambdafloweq}
\end{align}

\begin{figure}[t]
\normalsize
\begin{align*}
 \partial_k g_k= \ \tilde{\partial}_k &\left[ 
 \begin{tikzpicture}[baseline=-\the\dimexpr\fontdimen22\textfont2\relax]
\begin{feynhand}
\vertex[squaredot] (a) at (0,0){};
\vertex[squaredot](b) at (2,0){};
\vertex (c) at (-0.707,0.707){};
\vertex (d) at (2.707,.707){};
\vertex (e) at (-0.707,-0.707){};
\vertex (f) at (2.707,-.707){};
\graph{	(a)--[charged scalar, with arrow=0.5, half left,relative=false, out=90, in=90,looseness=1.2](b)--[anti fermion, half left, looseness=1.2](a), (c)--[charged scalar, with arrow=0.5] (a), (b)--[charged scalar, with arrow=0.5] (d),(e)--[ fermion] (a),(b)--[ fermion] (f)};
\end{feynhand}
\end{tikzpicture}+
 \begin{tikzpicture}[baseline=-\the\dimexpr\fontdimen22\textfont2\relax]
\begin{feynhand}
\vertex[squaredot] (a) at (0,0){};
\vertex[squaredot](b) at (2,0){};
\vertex (c) at (-0.707,0.707){};
\vertex (d) at (2.707,.707){};
\vertex (e) at (-0.707,-0.707){};
\vertex (f) at (2.707,-.707){};
\graph{	(a)--[charged scalar, with arrow=0.5, half left,relative=false, out=90, in=90,looseness=1.2](b)--[fermion, half left, looseness=1.2](a), (c)--[charged scalar, with arrow=0.5] (a), (b)--[charged scalar, with arrow=0.5] (d),(e)--[anti fermion] (a),(b)--[anti fermion] (f)};
\end{feynhand}
\end{tikzpicture} + 2
\begin{tikzpicture}[baseline=-\the\dimexpr\fontdimen22\textfont2\relax]
\begin{feynhand}
\vertex[squaredot] (a) at (-.75,0){};
\vertex[dot] (c) at (.75,.75){};
\vertex[dot](d) at (.75,-.75){};
\vertex (e) at (-1.457,0.707){};
\vertex (f) at (-1.457,-0.707){};
\vertex (g) at (1.457,1.457){};
\vertex (h) at (1.457,-1.457){};
\graph{(e)--[fermion, with arrow=0.5](a)--[charged scalar, with arrow=0.5,out=315,in = 180](d)--[fermion, with arrow=0.5](h), (f)--[charged scalar](a)--[fermion,out=45, in=180,  with arrow=0.5](c)--[charged scalar](g), (c)--[fermion](d) 	};
\end{feynhand}
\end{tikzpicture} + 2
\begin{tikzpicture}[baseline=-\the\dimexpr\fontdimen22\textfont2\relax]
\begin{feynhand}
\vertex[squaredot] (a) at (-.75,0){};
\vertex[dot] (c) at (.75,.75){};
\vertex[dot](d) at (.75,-.75){};
\vertex (e) at (-1.457,0.707){};
\vertex (f) at (-1.457,-0.707){};
\vertex (g) at (1.457,1.457){};
\vertex (h) at (1.457,-1.457){};
\graph{(e)--[fermion, with arrow=0.5](a)--[anti charged scalar,out=315,in = 180](d)--[fermion, with arrow=0.5](h), (f)--[anti charged scalar](a)--[fermion,out=45, in=180,  with arrow=0.5](c)--[anti charged scalar](g), (c)--[fermion](d) 	};
\end{feynhand}
\end{tikzpicture}
\right. \\
&+ 2 \left. \begin{tikzpicture}[baseline=-\the\dimexpr\fontdimen22\textfont2\relax]
\begin{feynhand}
\vertex[dot] (a) at (-.6,.6){};
\vertex[dot](b) at (-.6,.-.6){};
\vertex[dot] (c) at (.6,.6){};
\vertex[dot](d) at (.6,-.6){};
\vertex (e) at (-1.2803,1.2803){};
\vertex (f) at (-1.2803,-1.2803){};
\vertex (g) at (1.2803,1.2803){};
\vertex (h) at (1.2803,-1.2803){};
\graph{(f)--[anti fermion](b)--[anti fermion](d)--[anti charged scalar](h), (e)--[fermion](a)--[fermion, with arrow=0.5](c)--[charged scalar](g), (b)--[charged scalar](a), (c)--[fermion](d) 	};
\end{feynhand}
\end{tikzpicture}+ 2 \begin{tikzpicture}[baseline=-\the\dimexpr\fontdimen22\textfont2\relax]
\begin{feynhand}
\vertex[dot] (a) at (-.6,.6){};
\vertex[dot](b) at (-.6,.-.6){};
\vertex[dot] (c) at (.6,.6){};
\vertex[dot](d) at (.6,-.6){};
\vertex (e) at (-1.2803,1.2803){};
\vertex (f) at (-1.2803,-1.2803){};
\vertex (g) at (1.2803,1.2803){};
\vertex (h) at (1.2803,-1.2803){};
\graph{(f)--[anti fermion](b)--[anti fermion](d)--[anti charged scalar](h), (e)--[fermion](a)--[fermion, with arrow=0.5](c)--[charged scalar](g), (b)--[anti charged scalar](a), (c)--[fermion](d) 	};
\end{feynhand}
\end{tikzpicture} + 
\begin{tikzpicture}[baseline=-\the\dimexpr\fontdimen22\textfont2\relax]
\begin{feynhand}
\vertex[squaredot] (a) at (-.75,0){};
\vertex[dot] (c) at (.75,.75){};
\vertex[dot](d) at (.75,-.75){};
\vertex (e) at (-1.457,0.707){};
\vertex (f) at (-1.457,-0.707){};
\vertex (g) at (1.457,1.457){};
\vertex (h) at (1.457,-1.457){};
\graph{(e)--[charged scalar](a)--[anti fermion, out=315,in = 180](d)--[anti fermion](h), (f)--[anti charged scalar](a)--[fermion,out=45, in=180,  with arrow=0.5](c)--[fermion](g), (c)--[charged scalar](d) 	};
\end{feynhand}
\end{tikzpicture}  + 
\begin{tikzpicture}[baseline=-\the\dimexpr\fontdimen22\textfont2\relax]
\begin{feynhand}
\vertex[squaredot] (a) at (-.75,0){};
\vertex[dot] (c) at (.75,.75){};
\vertex[dot](d) at (.75,-.75){};
\vertex (e) at (-1.457,0.707){};
\vertex (f) at (-1.457,-0.707){};
\vertex (g) at (1.457,1.457){};
\vertex (h) at (1.457,-1.457){};
\graph{(e)--[charged scalar](a)--[anti fermion, out=315,in = 180](d)--[anti fermion](h), (f)--[anti charged scalar](a)--[fermion, out=45, in=180,  with arrow=0.5](c)--[fermion](g), (c)--[anti charged scalar](d) 	};
\end{feynhand}
\end{tikzpicture}
\right]
\\
\partial_k \lambda_k =\ \tilde{\partial}_k &\left[ 
 \begin{tikzpicture}[baseline=-\the\dimexpr\fontdimen22\textfont2\relax]
\begin{feynhand}
\vertex[squaredot] (a) at (0,0){};
\vertex[dot](b) at (2,0){};
\vertex (c) at (-0.707,0.707){};
\vertex (d) at (2.707,.707){};
\vertex (e) at (-0.707,-0.707){};
\vertex (f) at (2.707,-.707){};
\graph{	(a)--[charged scalar, with arrow=0.5, half left,relative=false, out=90, in=90,looseness=1.2](b)--[anti fermion, half left, looseness=1.2](a), (c)--[charged scalar, with arrow=0.5] (a),(e)--[ fermion] (a),(b)--[ fermion] (f)};
\end{feynhand}
\end{tikzpicture}+ \begin{tikzpicture}[baseline=-\the\dimexpr\fontdimen22\textfont2\relax]
\begin{feynhand}
\vertex[dot] (a) at (0,0){};
\vertex[squaredot](b) at (2,0){};
\vertex (c) at (-0.707,0.707){};
\vertex (d) at (2.707,.707){};
\vertex (e) at (-0.707,-0.707){};
\vertex (f) at (2.707,-.707){};
\graph{	(a)--[anti charged scalar, half left,relative=false, out=90, in=90,looseness=1.2](b)--[anti fermion, half left, looseness=1.2](a), (b)--[anti charged scalar] (d),(e)--[ fermion] (a),(b)--[ fermion] (f)};
\end{feynhand}
\end{tikzpicture}+
\begin{tikzpicture}[baseline=-\the\dimexpr\fontdimen22\textfont2\relax]
\begin{feynhand}
\vertex[dot] (a) at (-.75,0){};
\vertex[dot] (c) at (.75,.75){};
\vertex[dot](d) at (.75,-.75){};
\vertex (e) at (-1.457,0.707){};
\vertex (f) at (-1.457,-0.707){};
\vertex (g) at (1.457,1.457){};
\vertex (h) at (1.457,-1.457){};
\graph{(e)--[ fermion](a)--[charged scalar, with arrow=0.5,out=315,in = 180](d)--[fermion, with arrow=0.5](h),(a)--[fermion,out=45, in=180,  with arrow=0.5](c)--[anti charged scalar](g), (c)--[fermion](d) 	};
\end{feynhand}
\end{tikzpicture} +\begin{tikzpicture}[baseline=-\the\dimexpr\fontdimen22\textfont2\relax]
\begin{feynhand}
\vertex[dot] (a) at (-.75,0){};
\vertex[dot] (c) at (.75,.75){};
\vertex[dot](d) at (.75,-.75){};
\vertex (e) at (-1.457,0.707){};
\vertex (f) at (-1.457,-0.707){};
\vertex (g) at (1.457,1.457){};
\vertex (h) at (1.457,-1.457){};
\graph{(e)--[ fermion](a)--[anti charged scalar, out=315,in = 180](d)--[fermion, with arrow=0.5](h),(a)--[fermion, out=45, in=180,  with arrow=0.5](c)--[anti charged scalar](g), (c)--[fermion](d) 	};
\end{feynhand}
\end{tikzpicture}
\right]
\end{align*}
\caption{Diagrammatic representation of the flow equations of the coupling constants $g_k$ and $\lambda_k$. In the  flows of $\partial_k g_k$ (\cref{gfloweq}) and $\partial_k\lambda_k$ (\cref{lambdafloweq})  dashed lines denote exciton propagators and solid lines denote electron propagators. Dots denote the electron-exciton three point vertex ${\sim } \lambda_k$, while squares denote the electron-exciton four point vertex ${\sim} g_k$.}
\label{FeynmanFlow3}
\end{figure}
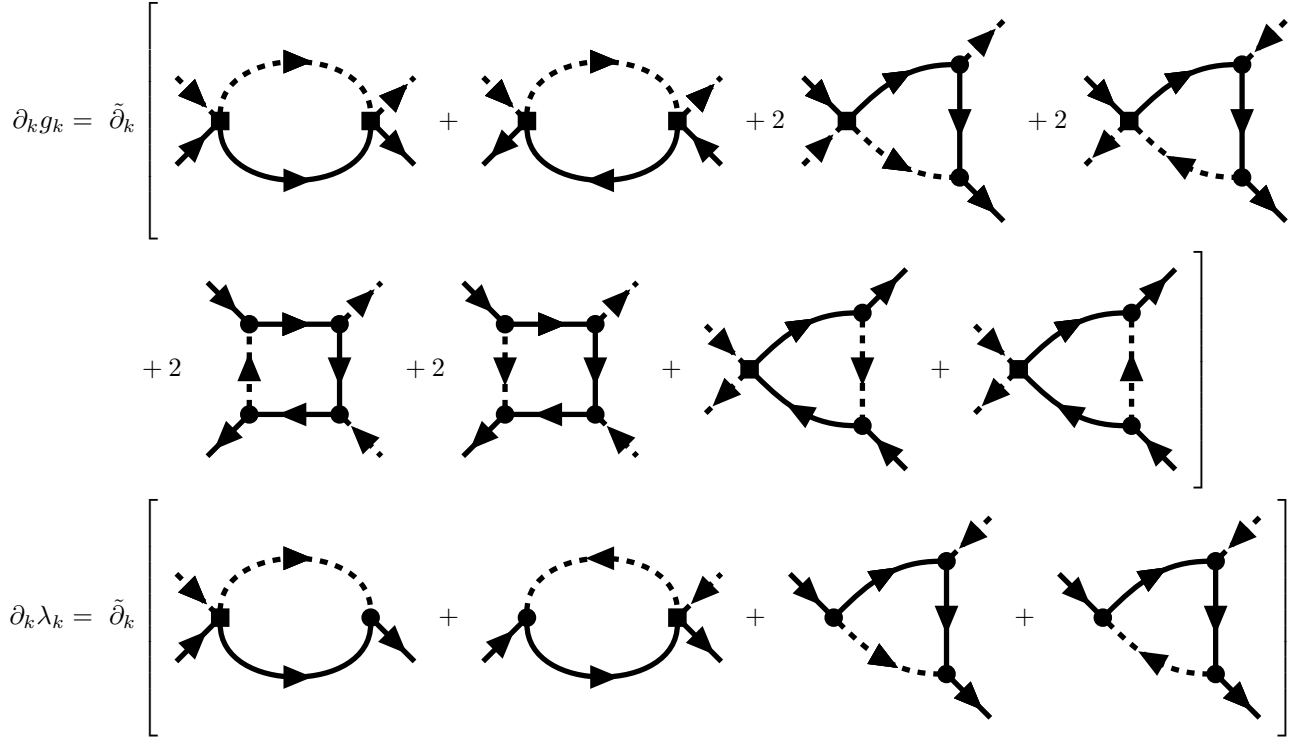

\begin{figure}[t]
\includegraphics[width=\linewidth]{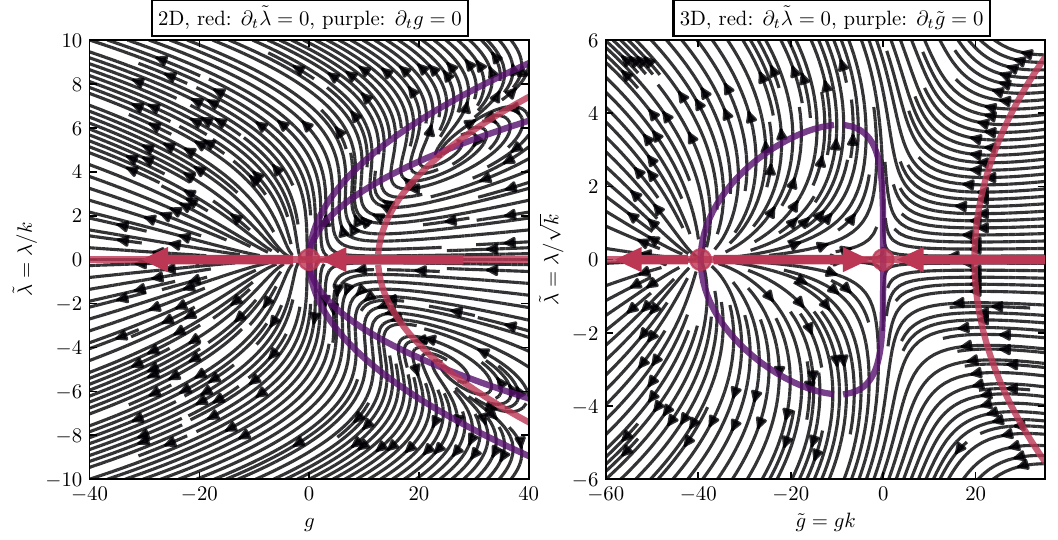}
\caption{Flow chart for the dimensionless coupling constants $\tilde \lambda$ and $\tilde g$ in (a) two dimensions and (b) three dimensions, obtained from a functional renormalization group analysis of the model \cref{truncation}. Using the dimensionless RG scale $t= \log(k/\Lambda)$, for given values of $g_k$ and  $\tilde{\lambda}_k= \lambda_k/k$   ( $\tilde{g}_k=g_k k$ and $\tilde{\lambda}_k= \lambda_k/ \sqrt{k}$ in 3D) the direction of the RG flow, given by the beta functions $\partial_t \tilde{\lambda}_t$ and $ \partial_t g_t $ ($\partial_t \tilde{\lambda}_t, \partial_t \tilde{g}_t $ in 3D) is shown. In the left panel the purple curves denote points with $\tilde{\lambda}_t^2=g_t$ and $\tilde{\lambda}_t^2= 2 g_t$ where $\partial_t g_t =0$ while the red curves denote points with $g_t= (8 \pi + \tilde{\lambda}_t^2) /2 $ and $\tilde{\lambda}_t=0$ where $\partial_t g=0$. Similarly, in the right panel  the purple curve denotes points with $\partial_t \tilde{g}_t=0$ and the red curves represent points with $\partial_t \tilde{\lambda}_t=0$. RG fixed points occur when the red and purple lines  cross.  The thick  arrows indicate the flow of $g$ and $\tilde{g}$ when $\tilde{\lambda}=0$.}
\label{flowchart}
\end{figure}
\subsection{Flow of coupling constants in two dimensions}

To evaluate the flow equations in the few-body limit, we set the chemical potentials  to $\mu_B=\mu_F=0$ following similar RG analysis, e.g., of the BEC-BCS crossover \cite{Nikolic2007}. After performing the momentum and frequency integrals, defining a dimensionless RG scale
\begin{align}
    t= \log\left(\frac{k}{\Lambda}\right)
\end{align}
and the dimensionless coupling constant  
\begin{align}
    \tilde{\lambda}_k= \frac{\lambda_k}{k},
\end{align}
 the  flow equations in dimensionless form read 
\begin{align}
    \partial_t g_t &= \frac{1}{2 \pi}\left( \frac{g_t^2}{2 }- \frac{3}{4}\tilde{\lambda}_t^2 g_t + \frac{\tilde{\lambda}_t^4}{4}\right) \\
    \partial_t \tilde{\lambda}_t&= \frac{1}{2 \pi}\left( \frac{\tilde{\lambda}_t g_t}{2 }- \frac{\tilde{\lambda}_t^3}{4}\right)- \tilde{\lambda}_t . 
\end{align}
The corresponding flow chart is shown in \cref{flowchart}, where flows begin in the UV at $t=0$ and end in the IR at $t=-\infty$.  For a given point in the flow diagram the arrows point in the direction of the flow. As one can see for $\tilde{\lambda}_{k=\Lambda}=0$ and $g_{k=\Lambda}>0$ the coupling constant $g$ flows towards the Gaussian, i.e. weak-coupling, fixed point, $\lim_{k\to 0} g_k=0$. On the other hand, for 
 $\tilde{\lambda}_{k=\Lambda}=0$ and $g_{k=\Lambda}<0$ the coupling flows to $\lim_{k\to 0} g_k= -\infty$, indicating bound state formation; this RG behavior reflects  that a bound state exists for any attractive interaction  in 2D \cite{Adhikari1986}. For $\tilde{\lambda}_{k=\Lambda}= \pm \sqrt{8 \pi}$, $g_{k=\Lambda}= \pm 8 \pi$ we find two additional repulsive fixed points, while for all other  initial values with $|\tilde{\lambda}_{k=\Lambda}|>0$ the flows are always driven towards $\lim_{k\to 0} g_k= -\infty$, meaning that bound state formation is inevitable and $g_k$ always represents a relevant correlation function that cannot be ignored. 

\subsection{Flow of coupling constants in three dimensions}
For completeness we also perform the RG analysis in three dimensions. After again using $t= \log(k/\Lambda)$ we can define the dimensionless coupling constants 
\begin{align}
    \tilde{g}_k= g_k k, \\
    \tilde{\lambda}_k= \frac{\lambda_k}{\sqrt{k}}
\end{align} and obtain the flow equations 
\begin{align}
    \partial_t \tilde g_t &= \frac{1}{2 \pi^2}\left( \frac{\tilde{g}_t^2}{2 }- \frac{3}{4}\tilde{\lambda}_t^2 \tilde{g}^{\phantom{2}}_t + \frac{\tilde{\lambda}_t^4}{4}\right) +\tilde{g}_t \\
    \partial_t \tilde{\lambda}_t&= \frac{1}{2 \pi^2}\left( \frac{\tilde{\lambda}_t \tilde{g}_t}{2 }- \frac{\tilde{\lambda}_t^3}{4}\right)-\frac{ \tilde{\lambda}_t}{2} . 
\end{align}
The resulting flow chart is shown in \cref{flowchart}(b). For  $\tilde{\lambda}_{k=\Lambda}=0$ it shows three different qualitative regions
\begin{align}
    \lim_{k\to 0 } \tilde{g}_k=  \left\{ \begin{aligned}
       &0^+,  &&\text{for } \tilde{g}_{k= \Lambda}>0 \\
       &0^-,  &&\text{for } - 4 \pi^2<\tilde{g}_{k= \Lambda}<0\\
       &-\infty, &&\text{for }   \tilde{g}_{k= \Lambda} <- 4 \pi^2 . 
    \end{aligned}  \right. 
\end{align} which yield different results with respect to the relevance of $\tilde{g}_k$.
For $\tilde{\lambda}_{k=\Lambda}\neq 0$, on the other hand, both dimensionless coupling constants are always relevant: 
\begin{align} 
&\lim_{k\to 0 } \tilde{g}_k= - \infty \\
&\lim_{k\to 0} \tilde{\lambda}_k=  \text{sign} (\tilde{\lambda}_{k=\Lambda} )\infty  
\end{align}
again demonstrating that there exists no scenario where bound state formation becomes irrelevant. Note, the fixed point at $\tilde g =-1$, $\tilde \lambda=0$ is the well-known fixed point representing the regime of unitary interactions in the BEC-BCS crossover in three dimensions.

\subsection{Discussion}

The flows of  coupling constants in \cref{flowchart} show a qualitatively similar picture in both two and three dimensions. Without the electron-exciton three-point vertex ${\sim}\lambda_k$ the relevance of the four-point vertex $g_k$ is dependent on the initial value of the four point vertex $g$.  In both cases, for repulsive initial values $g>0$ the four point vertex is irrelevant and vanishes as a result of the renormalization process $\lim_{k\to 0} g_k=0$. For attractive initial values $g<0$  in two dimensions the coupling is relevant and flows to strong-coupling physics featuring an exciton-electron bound state. In three dimensions, it is not sufficient that the coupling is attractive, but rather it needs to be sufficiently attractive $g<-4 \pi^2 / \Lambda$. If these conditions are fulfilled, the few-body system flows to strong coupling and thus the bound state physics needs to be taken into account. 

Considering the electron-exciton three-point vertex ${\sim}\lambda_k$, the qualitative nature of the relevance of the coupling changes. In two and three dimensions, apart from the two repulsive fixed points in 2D, a finite value of $\lambda$ always  leads to the system flowing to strong coupling, highlighting the relevance of the four-body vertex. This behaviour is akin to the \emph{in medium} behaviour of the two-body bound state in three dimensions: as the three-point vertex may be regarded as stemming from the immersion of a two-body problem within a bosonic medium $\lambda {\sim} \sqrt{n_0}g$, represented by the condensate. While in the vacuum two-body limit in three dimensions the bound state exists only for positive scattering lengths, when introducing a bosonic or fermionic medium, however, the two-body bound state exists for all scattering lengths. 

This analysis thus indicates that the four-point vertex is relevant and therefore needs to be taken into consideration, including the associated strong-coupling physics.

\section{Trion self-energy}

The trion self-energy $\Sigma_t^{\sigma}$ (see \cref{FigFeynCom}(c)) is given by
\begin{align}
\frac{\Sigma^{\sigma}_t(\pv, \omega)}{h^2}&=\lim_{T\to 0} \int_{\qv,n}G^{0}_{\psi^*_\sigma \psi^{\phantom{*}}_\sigma}(\pv-\qv,\omega-\nu_n) G^{0}_{\phi^* \phi^{\phantom{*}}} (\qv, \nu_n)  \nnl
&=\frac{1}{(2\pi)^3} \int d \qv d\nu \frac{1}{P_{\psi}(\pv-\qv, \omega-\nu) P_{\phi}(\qv,\nu)}.
\end{align} 
Here we have approximated  the diagram by its zero temperature $T=0$ expression which allows us to obtain an analytical result that can be readily employed in the following numerical computation. Based on favorable comparisons of $T=0$ theory and experimental observations at finite temperature in the Fermi polaron limit, we expect  finite temperature corrections to yield only small quantitative changes to the results. The microscopic short-range interaction has to be regularized and renormalized which gives the condition  \cite{Zoellner2011,Parish2011,vonMilczewski2022}
\begin{align}
    P_{t}^{0}(\qv,\omega)= \frac{h^2}{(2\pi)^2} \int_{|\qv|<\Lambda} d \qv \frac{1}{ \epsilon_T+2 \qv^2 },
\end{align}
where $\Lambda$ is the upper momentum cutoff \cite{Zoellner2011},
so that
\begin{align}
    P_t (\qv, \omega)= P_{t}^0 (\qv,\omega)- \Sigma^{\sigma}_t(\pv, \omega).
\end{align}
This function is related to the non-self-consistent $T$-matrix used commonly in single-channel approaches via $T(\pv, \omega)= - h^2 / P_t(\qv, \omega)$ \cite{vonMilczewski2022}.  For $\mu_F>0$ it is given in Eq. (3) of Ref \cite{Schmidt2012}. For $\mu_F<0$ and $\omega>0$ it is given by \cite{vonMilczewski2022}
\begin{align}
    P_t(\pv, \omega)= - h^2  \frac{i\pi + \log \left( \frac{\epsilon_T}{\mu_F +\mu_B - \pv^2 /2 + i\omega }\right)}{8\pi}, \ (-\mu_F, \omega >0)
\end{align}
and one can use $P_t(\pv, \omega)= P_t(\pv, -\omega)^*$ for $\omega<0$.

\section{Renormalized Green's functions}
Having introduced the trion self-energy to capture the strong coupling physics and the trion formation between electrons and excitons, the propagators used in the remaining diagrams are  computed using \cref{propagators} of the main text.
They are thus obtained as 
\begin{align}
    G_{\psi^*_{\sigma} \psi^{\phantom{*}}_{\sigma}} (\pv, \omega)&= \left(\frac{P_t}{P_t P_\psi - h^2 n_0}\right) (\pv, \omega)\\
    G_{t^*_{\sigma} t^{\phantom{*}}_{\sigma}} (\pv, \omega)&= \left(\frac{P_\psi}{P_t P_\psi - h^2 n_0}\right) (\pv, \omega)\\
    G_{t^*_{\sigma} \psi^{\phantom{*}}_{\sigma}} (\pv, \omega)&= \left(\frac{-h \sqrt{n_0}}{P_t P_\psi - h^2 n_0}\right) (\pv, \omega)\\
    G_{\psi^*_{\sigma} t^{\phantom{*}}_{\sigma}} (\pv, \omega)&= \left(\frac{-h \sqrt{n_0}}{P_t P_\psi - h^2 n_0}\right) (\pv, \omega),
\end{align}
 and $ G_{\phi^* \phi^{\phantom{*}}}(\pv, \omega)= 1/ P_\phi (\pv, \omega)$,  where changing the order of fermionic (bosonic) indices results in an additional factor of $-1$ (1). The remaining matrix elements of the matrix valued Green's function $G$ vanish. 

 Similarly, the matrix elements of the bare Green's function $G_0$ can be computed where greater care of needs to be taken with respect to the upper cut-off scale $\Lambda$. This comes about as for $\Lambda\to \infty$ we have that $g=-h^2/P^{0}_t\to 0$, while $-h^2/(P_t^0-\Sigma^\sigma_t )=-h^2/P_t\quad  \xcancel{\to }\quad 0 $. As a result, to avoid that cancelling $\Lambda$ dependencies are overlooked, the limit $\Lambda\to \infty$ may not be taken prematurely  
 \begin{align}
    G^0_{\psi^*_{\sigma} \psi^{\phantom{*}}_{\sigma}} (\pv, \omega)&= \left(\frac{P^0_t}{P^0_t P_\psi - h^2 n_0}\right) (\pv, \omega)\xrightarrow{\Lambda\to \infty} \left(\frac{1}{ P_\psi }\right) (\pv, \omega)\\
    G^0_{t^*_{\sigma} t^{\phantom{*}}_{\sigma}} (\pv, \omega)&= \left(\frac{P_\psi}{P^0_t P_\psi - h^2 n_0}\right) (\pv, \omega)\xrightarrow{\Lambda\to \infty} \left(\frac{1}{ P^0_t }\right) (\pv, \omega)\label{G0}.
\end{align}

\section{Exciton self-energy}
As discussed in the main text, the bosonic chemical potential is fixed by the Hugenholtz-Pines relation used in the condition (ii) of the main text. The exciton self-energy entering this condition is represented by the diagram in \cref{FigFeynCom}(g).  It is given by
\begin{align}
\Sigma_B(\qv,\nu_m)=  \frac{h^2 T}{(2\pi)^2} \int d \pv \sum_{n,\sigma} &G_{t^*_\sigma t^{\phantom{*}}_\sigma} (\pv+\qv, \omega_n+\nu_m) G_{\psi^{\phantom{*}}_\sigma  \psi^*_\sigma  } (\pv, \omega_n).
\end{align}
Instead of numerically evaluating the Matsubara sum directly (leading to poor convergence), we rather compute an equivalent contour integral for which a contour is laid around the Matsubara frequencies and then deformed to the real axis to arrive at
\begin{align}
    \Sigma_B(0,0)&=  \sum_{\sigma} \frac{1}{4 \pi^3} \int d \pv\int_{-\infty}^{\infty} d\Omega n_F(\Omega) \Im \left(\frac{P_t(\pv, - i z) P_\psi (\pv,-iz)}{\left[P_t(\pv, -iz) P_\psi (\pv,-iz)- h^2 n_0\right]^2} \right)\Bigg|_{z= \Omega + i \epsilon}
\end{align}
where $n_F(\Omega)= 1/(1+ e^{\Omega/T})$ is the Fermi-distribution function. This allows for efficient numerical evaluation.

At $T=0$ and for $n_0=0$, the Hugenholtz-Pines relation (ii) determines the energy of Fermi polarons in the limit of a single impurity in a bath of fermions. This is equivalent to the treatment carried out in common variational and non-selfconsistent approaches \cite{Chevy2006,Combescot2007,Punk_2009,Zoellner2011,Trefzger_2012,Schmidt2012}. To see that, one may notice that for $n_0=0$ the dressed fermionic propagator $G_{\psi^{\phantom{*}}_{\sigma} \psi^*_{\sigma}}$ and the dressed trion propagator $G_{t^*_\sigma t^{\phantom{*}}_\sigma}$ within $\Sigma_B$ reduce to the bare fermionic propagator $G^0_{\psi^{\phantom{*}}_{\sigma} \psi^*_{\sigma}}$ and and $1/P_t$, respectively. After identifying the $T$-matrix as $T=-h^2  G_{t^*_{\sigma} t^{\phantom{*}}_{\sigma}} =- h^2 / P_t$, one finds equivalent diagrams and the Hugenholtz-Pines condition precisely gives the condition required to find the energy of the Fermi polaron.

\section{Fermion number equation}
Similar to the exciton self-energy $\Sigma_B$, the Matsubara summation for the number equation, entering the condition (\romannumeral 1) of the main text, converges only slowly. Hence we again deform the integration contour  to wrap around the real axis. In this way, the fermion density can be computed as follows:
\begin{align}
    n_F=&-  \frac{1}{(2\pi)^2} \int d \pv \frac{1}{\beta}\sum_n  \frac{P_t(\pv, \omega_n) }{P_t(\pv, \omega_n) P_\psi (\pv, \omega_n)- h^2 n_0}\nnl
        =&  \frac{1}{(2\pi)^2} \int d \pv \frac{1}{ \pi}\int_{-\infty}^{\infty} d\Omega n_F(\Omega)   \Im \left(\frac{P_t(\pv, - i z) }{P_t(\pv, -iz) P_\psi (\pv,-iz)- h^2 n_0} \right)\Bigg|_{z= \Omega + i \epsilon} \label{numberequation}
\end{align}

At $T=0$, the critical point where $n_F=0$ turns to $n_F>0$ determines the energy of the Bose polaron (a single impurity in a bath of condensed bosons), equivalent to the treatment in Ref. \cite{rath2013}. This can be seen by noting that for $n_F=0$ the Hugenholtz-Pines condition (ii) imposes a vanishing boson chemical potential and identifying the fermionic self-energy as $h^2n_0/ P_t$ within the fermionic propagator $\sim 1/(P_\psi - h^2 n_0 /P_t)$. This leads to the same diagrams and expressions and the critical point for $n_F=0$ then corresponds to the condition needed to solve for the energy of the Bose polaron.

\section{Computation of electron-trion scattering vertex}

We perform an $s$-wave projection of the electron-trion scattering vertex  in which we consider scattering at the Fermi-wavevector $k_F$ of the balanced two component Fermi gas of electrons
\begin{align}
 \tilde{\gamma}_{\sigma, \sigma'}= \frac{1}{2 \pi} \int d\theta_{\kv,\kv'} \frac{1}{2}  \Big[& \gamma(K_+-K_+',K_+,K_+')_{\sigma,\sigma'}  +\gamma(K_- -K_-',K_-,K_-')_{\sigma,\sigma'}\Big]. \label{lambdaswave}
\end{align}
Here $K_\pm=(\kv, \pm\pi T),K_\pm'=(\kv', \pm \pi T)$, $|\kv|=|\kv'|=k_F$, $\theta_{\kv,\kv'}$ denotes the angle between $\kv$ and $\kv'$ and $\tilde{\gamma}_{\sigma, \sigma'}$ is used within \cref{BSE}.

Using \cref{BSE,lambdaswave} the expression for the $s$-wave projection of the electron-electron scattering vertex is then given by
\begin{align}
\tilde{\gamma}_{\sigma,\sigma'}= \tilde{\gamma}^0_{\sigma,\sigma'}+  \tilde{\gamma}_{\sigma,\sigma'}\int_{\pv}&\frac{T}{2}\sum_{\omega_n}\left(\frac{1}{P_\phi (\pv-\kv,\omega_n+ \pi T)}+\frac{1}{P_\phi (\pv-\kv,\omega_n-\pi T)}\right)\nnl
&\quad \times \frac{h^2 n_0}{\left[P_t(\pv, \omega_n) P_\psi (\pv, \omega_n)- h^2 n_0\right] \left[P_t(-\pv,-\omega_n) P_\psi (-\pv,-\omega_n)- h^2 n_0\right]}, \label{lambdatildeexpression}
\end{align}
where $|\kv|=k_F$. The pairing instability is computed by rearranging \cref{lambdatildeexpression} to $\tilde{\gamma}_{\sigma,\sigma'}= \tilde{\gamma}^0_{\sigma,\sigma'} /(1- F) $ and solving for $F=1$ where 
\begin{align}
    F=\frac{ \tilde{\gamma}_{\sigma,\sigma'}-\tilde{\gamma}^0_{\sigma,\sigma'}}{ \tilde{\gamma}_{\sigma,\sigma'}}.
\end{align}
 $F$ represents the integral and sum part in \cref{lambdatildeexpression}. As this integral decays  faster in frequency than the number equation and the exciton self-energy, the Matsubara summation in \cref{lambdatildeexpression} can be directly computed numerically, without the need to deform the integration contour.

\begin{figure*}[t]
\includegraphics[width=\linewidth]{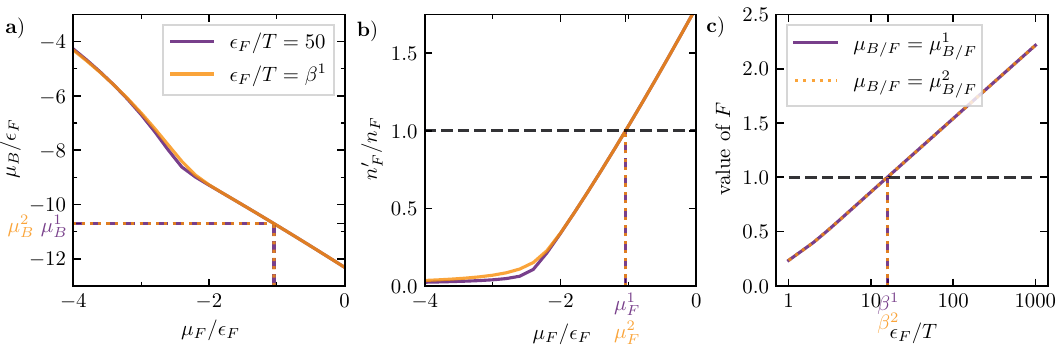}
\caption{Exemplary optimization procedure to simultaneously solve for the Hugenholtz-Pines relation and the number equation at $\epsilon_T=5 \epsilon_F $, $n_0/n_F=1$. (a) The  boson chemical potential $\mu_B(\mu_F,n_0, \epsilon_T,T)$ to fulfill the Hugenholtz-Pines relation is shown as a function of the fermion chemical potential. At an initial  temperature of $T_0=\epsilon_F/50 $ (purple) the critical boson chemical potential is computed as a function of the fermion chemical potential and the resulting pairs of ($\mu_F, \mu_B(\mu_F,n_0, \epsilon_T,T_0)$) are used to compute the corresponding fermion density $n'_F(\mu_F, \mu_B(\mu_F,n_0, \epsilon_T, T_0))$, shown in (b). The point at which $n'_F(\mu^1_F, \mu_B(\mu^1_F,n_0, \epsilon_T, T_0))= n_F$ is used to determine $\mu_F^1$. From this, the corresponding boson chemical potential $\mu_B^1$ is computed as  $\mu_B^1 = \mu_B(\mu^1_F,n_0, \epsilon_T,T_0)$. In (c) the value of F obtained using $\mu_F^1$ and $\mu_B^1$ is shown  for varying temperatures and it is used to determine a critical inverse temperature $\beta^1=1/T_1$ by locating where $F=1$. This temperature is used in a second iteration (yellow)  from which  $\mu_F^2$ and $\mu_B^2$ are found which are used in  (c) to find a critical inverse temperature $\beta^2=1/T_2$. This cycle is repeated until the  chemical potentials and critical temperatures are found to be converged.  }
\label{muFinder}
\end{figure*}

\section{Determining the critical pairing temperature}

To estimate the critical pairing temperature for given values of $\epsilon_T/ \epsilon_F$ and $n_0/n_F$, the critical pairing condition  needs to be solved for,  while fulfilling the number equation (\romannumeral 1) and the Hugenholtz-Pines relation (\romannumeral 2). This is done in a self-consistent optimization procedure which we describe in the following. 

First, for given values of $n_0/\epsilon_F, \epsilon_T/\epsilon_F$ and an initial temperature of $T_0=\epsilon_F/50$  the critical boson chemical potential to fulfill the Hugenholtz-Pines relation (\romannumeral 1) is computed as $\mu_B(\mu_F,n_0, \epsilon_T,T_0)$ for a varying fermion chemical potential $\mu_F$. Next, these chemical potentials are used within the number equation \cref{numberequation} to compute the Fermi density $n'_F(\mu_F, \mu_B(\mu_F,n_0, \epsilon_T, T_0))$. From this, the fermion chemical potential $\mu_F^1$ fulfilling $n_F=n'_F(\mu_F^1, \mu_B(\mu_F^1,n_0, \epsilon_T, T_0))$ is found and the corresponding boson chemical potential is determined as $\mu_B^1= \mu_B(\mu_F^1,n_0, \epsilon_T,T_0)$. 

Using $\mu_F^1$ and $\mu_B^1$, the critical temperature $T_1$ where $F=1$ is then found using \cref{lambdatildeexpression}. This critical temperature $T_1$ is then used as an input to find $\mu_F^2 $ and $\mu_B^2$ which are in turn used to find a critical temperature $T_2$. This cycle is repeated until the chemical potentials and the temperature have converged to a fixed point which simultaneously satisfies the number equation, the Hugenholtz-Pines relation and the critical pairing condition. For given values of $n_0/\epsilon_F, \epsilon_T/\epsilon_F$ the  temperature found gives the critical pairing temperature $T_c^*$. This procedure is shown in \cref{muFinder} for the first two iterations of this cycle.

\begin{figure}[b]
\centering
\includegraphics[width=0.7\linewidth]{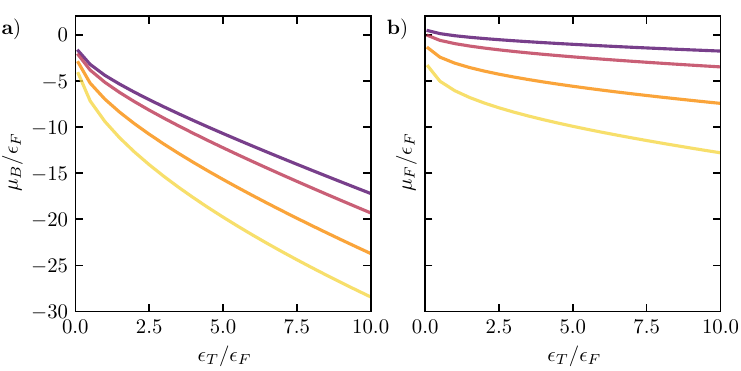}
\caption{The boson (a) and fermion (b) chemical potentials $\mu_B$ and $\mu_F$ at the critical temperature $T_c^*$ are shown as a function of $\epsilon_T/\epsilon_F$ for the condensate densities $n_0/n_F=1$ (purple), $2$ (red), $5$ (orange), and $10$ (yellow). These potentials are simultaneous solutions of the number equation (i) and the Hugenholtz-Pines relation (ii), and to satisfy these, the chemical potentials increase in magnitude with increasing binding energy.}
\label{chemicalpotentialsatcritpairing}
\end{figure}
The resulting chemical potentials for the results shown in \cref{crittemp} are given in \cref{chemicalpotentialsatcritpairing}.

\section{Determining the bipolaron binding energy and the boundary of the BCS regime}
As discussed in the main text, the method used to obtain the critical pairing temperature provides a reasonable approximation for the critical temperature of superfluidity in the regime where a BCS-type theory is appropriate. The $s$-wave projected pairing vertex $\tilde{\gamma}_{\sigma, \sigma'}$ defined in \cref{lambdaswave} admits a bound state at $T=0$ even in the limit where the Fermi density vanishes, which we refer to as the polaron limit. Thus finding a singularity in $\tilde \gamma$ implies the formation of a bound state between two Bose polarons, a bound state which  we refer to as the bipolaron \cite{Camacho2018}. In the strong coupling limit, we expect the superfluid transition temperature $T_c$ to be more accurately captured by a BKT theory of a Bose gas of bipolarons. As a result we approximate the point where the system crosses over from a BCS-type to a BKT/BEC-type  behaviour as the point where the bipolaron binding energy becomes comparable to the Fermi energy and, as a result, the binding length of the bipolaron is comparable to the average fermion interparticle distance.  

At $T=0$ in the polaron limit ($n_F=0$) the exciton self-energy vanishes identically and as a result we set $\mu_B=0$. Thus for given values of $\epsilon_T$, $n_0$ there exists a critical chemical potential $\mu_{F,n_F=0} ( n_0, 
\epsilon_T)$ for which $n_F=0 $  for $\mu_F<\mu_{F,n_F=0} ( n_0, \epsilon_T)$ and $n_F>0 $ for $\mu_F>\mu_{F,n_F=0} ( n_0, 
\epsilon_T)$. This chemical potential in fact determines the Bose polaron energy which, for three dimensional systems, has been shown to agree remarkable well with experimental observations \cite{Hu2016}.

The binding energy of the bipolaron is determined from the divergence of $\tilde{\gamma}_{BP}$, 
\begin{align}
\tilde{\gamma}_{BP}&= \tilde{\gamma}^0_{BP}+  \tilde{\gamma}_{BP}\int_{\pv}\int \frac{d \omega}{2 \pi}\frac{ h^2 n_0 }{P_\phi (\pv,\omega)}  \frac{1}{P_t(\pv, \omega) P_\psi (\pv, \omega)- h^2 n_0}   \frac{1}{P_t(-\pv, -\omega) P_\psi (-\pv, -\omega)- h^2 n_0}, \label{bipolaronvertex}
\end{align}
which is obtained from \cref{lambdatildeexpression} in the limit $k_F\to 0$, $T \to 0$.
The divergence of $\tilde{\gamma}_{BP}$ occurs at a fermion chemical potential $\mu_{F, BP} (n_0, \epsilon_T)< \mu_{F,n_F=0}( n_0, 
\epsilon_T)$.  Thus the  bipolaron binding energy is given as
\begin{align}
    E_{BP}= 2 \left(\mu_{F, BP} (n_0, \epsilon_T)- \mu_{F,n_F=0}( n_0, 
\epsilon_T)\right).
\end{align}

\begin{figure}[t]
\includegraphics[width=0.7\linewidth]{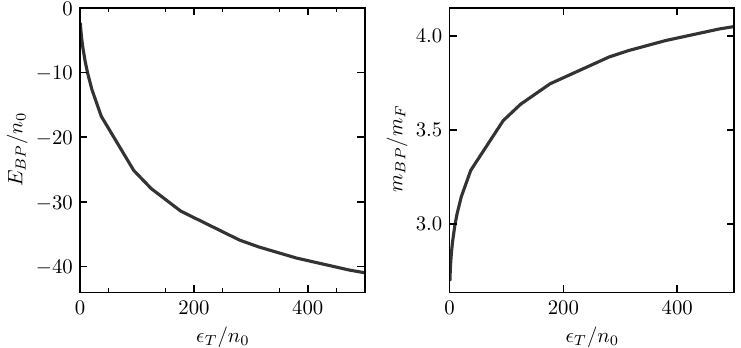}
\caption{Bipolaron binding energy $E_{BP}$ (left) and bipolaron effective mass $m_{BP}$ (right) as a function of the trion energy $\epsilon_T$. The energies are given in units of condensate density $n_0$ while the mass is given in units of the fermion mass $m_F$. With increasing $\epsilon_T$ the bipolaron becomes deeper bound and acquires a moderate effective mass.}
\label{BPbindingenergy}
\end{figure}

The resulting bipolaron binding energies are shown in \cref{BPbindingenergy}. Hence, requiring  the binding energy per particle of the bipolaron to be smaller than the Fermi energy each fermion experiences, we require  $ \mu_{F,n_F=0}( n_0, 
\epsilon_T) - \epsilon_F <\mu_{F, BP} (n_0, \epsilon_T)< \mu_{F,n_F=0}( n_0, 
\epsilon_T)$ for the BCS theory to be applicable. The resulting critical dimensionless interaction strengths $\epsilon_T/\epsilon_F$  for given values of $n_0/n_F$ are shown in \cref{bcsboundary} of the main text and the end points of the BCS regime are indicated in \cref{crittemp} of the main text. 

\section{Approximation of the BKT transition temperature}
To approximate the critical temperature for the transition into a BKT superfluid, we use \cite{Fisher1988,Proko2001,Proko2002,Petrov2003}
\begin{align}
    T_{BKT} = \frac{2 \pi n_F}{m_{BP}} \frac{1}{ \log \left(\frac{\eta}{4 \pi} \log \left(\frac{1}{n_F d_*^2}\right) \right)}, \label{BKTtemp1}
\end{align}
where the density of bipolarons is given by $n_F$ (all fermions can be assumed to be paired into bipolarons), and $\eta \approx 380$ \cite{Petrov2003}.  The bipolaron-bipolaron scattering length is given by $d_*$ , and $m_{BP}$ is the effective bipolaron mass. The bipolaron scattering length is approximated by the binding length of the bipolaron \cite{Petrov2003}, which in turn is parametrized by the bipolaron binding energy as 
\begin{align}
    d_* = \sqrt{-\frac{1}{2 E_{BP}} \left(\frac{1}{m_F}+ \frac{1}{m_F+m_B}  \right)} . 
\end{align}
The bipolaron effective mass is computed by evaluating \cref{bipolaronvertex} at a finite incoming momentum $\qv$ which is distributed along the two fermionic propagator legs as $\pv+ \qv/2$ and $\qv/2- \pv$. From this, the bipolaron dispersion relation is computed as a function of $|\qv|$ and the effective bipolaron mass $m_{BP}$ is obtained from a quadratic  fit to this dispersion relation. The resulting bipolaron effective mass is shown in \cref{BPbindingenergy}. The BKT transition temperatures obtained from \cref{BKTtemp} are shown in \cref{crittemp} of the main text.

\end{widetext}

\end{document}